\documentclass[prl,aps,twocolumn,epsfig,nofootinbib,floats]{revtex4}

\usepackage{amsmath}
\usepackage{graphics}
\usepackage{epsfig}
\begin{document}
\preprint{\vbox{\hbox{UCI-TR-2008-16, NSF-KITP-08-57}}}
\title{\Large\bf Minimal Flavor Violation in the Lepton Sector of the Randall-Sundrum Model}
\author{\bf Mu-Chun Chen$^{1,2,}$\footnote{e-mail: muchunc@uci.edu} and Hai-Bo Yu$^{1, }$\footnote{
  e-mail: haiboy@uci.edu}}

\affiliation{$^{1}$Department of Physics \& Astronomy, University of California, Irvine, CA 92697, USA\\
$^{2}$KITP, University of California, Santa Barbara, CA 93107, USA}

\date{April, 2008}
\begin{abstract}
We propose a realization of Minimal Flavor Violation
 in the lepton sector of the Randall-Sundrum model. With the MFV
assumption, the only source of flavor violation are the 5D Yukawa
couplings, and the usual two independent sources of flavor violation
are related. In the limit of massless neutrinos, the bulk mass
matrices and 5D Yukawa matrices are simultaneously diagonalized, and
hence the absence of FCNCs. In the case of massive neutrinos, the
contributions to FCNCs in the charged lepton sector are highly
suppressed, due to the smallness of neutrino masses. In addition, the MFV assumption also allows suppressing one-loop charged current contributions to flavor changing processes by reducing the size of the Yukawa couplings, which is not possible in the generic anarchical case. We found that the first KK mass scale as low as $\sim 3$ TeV can be allowed. In
both cases, we present a set of numerical results that give rise to realistic lepton masses and mixing angles. Mild hierarchy in the 5D Yukawa matrix of $\mathcal{O}(25)$ in our numerical example is required to be consistent with two large and one small mixing angles. This tuning could be improved by having a more thorough search of the parameter space.

\end{abstract}
\maketitle

\section{Introduction}

The Randall-Sundrum (RS) model~\cite{Randall:1999ee} is a solution
to the gauge hierarchy problem based on non-factorizable geometry in
a slice of $AdS_{5}$ space with warped background metric. By
allowing the fermion and gauge fields to propagate in the bulk
\cite{bulkfields}, the model can accommodate fermion mass hierarchy
by localizing different fermions at different locations along the
fifth dimension, while having all 5D Yukawa couplings to be ${\cal
O}(1)$~\cite{bulkfermion1,bulkfermion2}. It also gives rise to novel
ways to generate small neutrino
masses~\cite{bulkfermion1,Chen:2005mz}.
Realistic models based on bulk custodial
symmetry \cite{custodial} or large brane kinetic terms
\cite{branekinetic} have also been built, in which the first KK mass scale $\sim 3~{\rm TeV}$ is allowed by
the electroweak precision data. The RS model with bulk fermions and gauge bosons
has a rich flavor structure. There are two sources of flavor
violation, the 5D Yukawa couplings and the bulk fermion mass terms,
which can generate dangerous tree-level FCNCs mediated by
the KK gauge bosons. Even though these processes are somewhat
suppressed by the built-in RS GIM mechanism~\cite{rsgim1},
constraints from the CP-violating parameter $\epsilon_K$ for
$K^{o} - \overline{K^{o}}$ mixing still gives a stringent bound of 8 TeV with a
generic flavor structure~\cite{Bona:2007vi}. A detailed scanning of
the parameter space has shown recently that the first KK gluon mass
should be heavier than about 21 TeV for the generic anarchy
case~\cite{Csaki:2008zd}.

Lepton flavor violation (LFV)  in various rare leptonic processes
mediated by neutral KK gauge bosons also gives stringent constraints
on the KK mass scale. LFV in the RS model has been studied
before~\cite{kitano,lfv,lfvagashe}. Even in the case of massless
neutrinos, severe bound on the first KK mass scale already arises
from FCNC-mediated processes in scenarios with generic anarchical 5D
Yukawa couplings~\cite{lfvagashe}. Moreover, there is a tension
between loop induced processes, such as $\mu\rightarrow e\gamma$,
and the tree-level flavor violating processes, such as $\mu-e$
conversion, since they depend on the 5D Yukawa coupling constants
oppositely~\cite{lfvagashe}. As a result, the allowed parameter
space for the 5D Yukawa couplings is very restricted and it is not
possible, in the generic anarchy case, to relax the bound on the KK mass scale by tuning the 5D
Yukawa coupling constants.

Suppression of flavor violation with bulk and brane flavor symmetries
has been studied~\cite{Cacciapaglia:2007fw}. In \cite{rsmfv},
the assumption of minimal flavor violation (MFV)~\cite{mfv} is 
applied to the quark sector of the RS model, which assumes
that the 5D Yukawa couplings are the only source that breaks the global flavor symmetry and
 the bulk mass matrices are related to the 5D Yukawa couplings as dictated by the
flavor symmetry.  This thus provides an alignment between the two flavor violating sources
that are otherwise independent in the general
case.

In this Letter, we propose a realization of MFV in the lepton sector
of the RS model. In the massless neutrino case, the 5D bulk mass
matrices and 5D Yukawa matrices can be diagonal simultaneously.
There are thus no leptonic FCNCs at tree level. In the presence of massive Dirac
neutrinos, due to the MFV assumption, the FCNC contributions are
controlled by one single parameter, $\xi$. Interestingly, small
neutrino masses requires $\xi \lesssim 0.1$, leading to very
suppressed FCNCs and thus allowing a light KK mass scale, making it
possible to test the model at the collider~\cite{collider}. Furthermore, MFV also alleviates the tension that exists in the generic anarchical case between the tree-level FCNCs and one-loop charged-current contributions to LFV processes. It allows the charged-current contributions to be suppressed by reducing the Yukawa couplings without increasing the tree-level FCNC contributions.

\section{Minimal Flavor Violation in the Lepton Sector}

{\it Massless Neutrino Limit.}
With the MFV assumption, all flavor violation come from the 5D charged lepton Yukawa couplings $Y_e$.
The relevant bulk mass matrices and the 5D Yukawa couplings are
\begin{eqnarray}\label{L1}
{\cal L}_{\rm 5D}^{\rm lep} \supset \overline{L}C_LL+\overline{e}C_e e + \overline{H} \, \overline{L} Y_e e \; .
\end{eqnarray}
To implement the MFV in the lepton sector, we assume the bulk mass
matrices for the lepton doublets and singlets are aligned with the 5D Yukawa coupling
$Y_e$ as
\begin{eqnarray}
C_L= b Y_e Y^{\dagger}_e, \quad  C_e = a Y^\dagger_e Y_e  \; ,
\label{eq:mfv-mlnu}
\end{eqnarray}
where $a$ and $b$ are ${\cal O}(1)$ parameters\footnote{For
simplicity, we omit flavor symmetry invariant identity contributions
to the 5D mass parameters. Including these terms does not change our
conclusions.}. With $U(3)_L\times U(3)_e$ global flavor symmetry, we
can select a basis such that $Y_e$ is in the diagonal form,
$\hat{Y}_e$. In this basis, both bulk mass matrices $C_{e} = a
\hat{Y}_{e}^{\dagger} \hat{Y}_{e}$ and
$C_L=b\hat{Y}_e\hat{Y}^{\dagger}_e$ are diagonal and thus all
leptonic flavor violation vanish. 
We comment that in the Randall-Sundrum framework, the Yukawa couplings can in general be of order $\mathcal{O}(1)$. In that case, there are large higher order corrections involving four and more Yukawa matrices to Eq.~(\ref{eq:mfv-mlnu}) in the expansion. Even though these higher order corrections will change the numerical fits to $\hat{Y}_{e}$, with massless neutrinos, the matrices $C_{L}$ and $C_{e}$ can still be diagonalized simultaneously. As a result, the tree-level FCNCs are still absent. From the numerical solutions given below, one sees that in order to obtain realistic lepton masses, the Yukawa couplings are all small compared to $1$, and thus the expansion in Eq.~(\ref{eq:mfv-mlnu}) is justified.

We next show that the MFV assumption in Eq.~(\ref{eq:mfv-mlnu}) can
indeed give rise to realistic charged lepton masses. In the brane
Higgs case\footnote{
In our numerical study, we assume the brane Higgs limit.
Our results can be easily extended to the bulk Higgs case~
\cite{bulkhiggs}.}, the charged lepton masses are given by
$m_l \simeq v F_L Y_e F_e$,
where $v=174$ GeV, and $F_L$ and $F_e$ are the values of the zero-mode profiles on the TeV brane for
the lepton doublets and singlets, respectively. The eigenvalues of $F_{L}$ and $F_{e}$ are given by
\begin{eqnarray}
f_{L_i}=\sqrt{\frac{1-2c_{L_i}}{1-\epsilon^{1-2c_{L_i}}}}, \; f_{e_i}=\sqrt{\frac{1-2c_{e_i}}{1-\epsilon^{1-2c_{e_i}}}}
\end{eqnarray}
where $\epsilon = e^{-\pi k r_{c}}\simeq10^{-15}$ and $c_{L_i}$ and
$c_{e_i}$ are the eigenvalues of the 5D bulk mass matrices $C_L$ and
$C_e$. As shown above, the matrices $Y_e$, $C_L$ and $C_e$ can be
diagonal simultaneously due to the MFV assumption. In the diagonal basis, 
$Y_e={\rm diag}(Y_{e_1}, \, Y_{e_2}, \, Y_{e_3})$, $C_L={\rm
diag}(b|Y_{e_1} |^2, \, b|Y_{e_2}|^2, \, b|Y_{e_3}|^2)$ and
$C_e={\rm diag}(a|Y_{e_1}|^2, \, a|Y_{e_2}|^2, \, a|Y_{e_3}|^2)$.
Without loss of generality, we choose $a=1$ and $b=1$. With the
values of $Y_{e_1}\simeq 0.816, \; Y_{e_2}\simeq 0.759$ and 
$Y_{e_3}\simeq 0.720$, the following realistic charged lepton masses
are obtained, $m_e \simeq 0.511~{\rm MeV}, \; m_\mu\simeq105.6~{\rm
MeV}$ and $m_\tau\simeq1.77~{\rm GeV}$. We note that even if $a \ne b$, the matrices $C_{L}$, $C_{e}$ and $Y_{e}$ can all still be diagonalized simultaneously and thus avoid the tree-level FCNCs.

{\it Massive Neutrino Case.} To accommodate massive neutrinos and
lepton mixing, we introduce three right-handed (RH) neutrinos  in the
model. The RH neutrinos reside in different $SU(2)_{R}$ doublets
\cite{custodial} from those that contain the iso-spin singlet
charged leptons. The RH neutrinos couple to the lepton doublets to
form Dirac mass terms through the Yukawa coupling $Y_{\nu}$. The
relevant Lagrangian in this case  is given by\footnote{Here we implicitly assume lepton number conservation so that no Majorana mass term $LLHH$ is generated.}   
\begin{eqnarray}\label{L2}
{\cal L}_{\rm 5D}^{\rm lep}  \supset
\overline{L}  C_LL+\overline{e}C_e e +\overline{N}C_NN+\overline{H} \, \overline{L}Y_e e + H \overline{L}Y_\nu
N \; .
\end{eqnarray}
The smallness of neutrino masses is then archived by localizing the
RH neutrinos close to the Planck brane such that their overlap with
the lepton doublets is small.

With the MFV assumption, the 5D bulk mass matrices are related to
the 5D Yukawa couplings as
\begin{eqnarray}
C_e= aY^\dagger_eY_e,~~C_N=dY^\dagger_\nu Y_\nu,~~C_L=c(\xi Y_\nu
Y^\dagger_\nu+Y_e Y^\dagger_e) \; ,
\label{eq:mfv2}
\end{eqnarray}
where $a,~d,~c$ are ${\cal O}(1)$ parameters and $C_{N}$ is the bulk
mass term for the RH neutrinos. 
With three RH neutrinos, the global
flavor symmetry is $U(3)_L\times U(3)_e \times U(3)_N$, with which
one can rotate to a basis where either $Y_e$ or $Y_\nu$ is diagonal.
In the following analysis, we work in the basis in which $Y_e$ is
diagonal and it is denoted by $\hat{Y}_e$. In this basis, $Y_\nu$
can be written as  $Y_\nu=V_{5D}\hat{Y}_\nu$, where $V_{5D}$ is the
5D leptonic mixing matrix. All flavor mixing in the lepton sector
are generated by $V_{5D}$. In this basis, both $C_e$ and $C_N$ are
diagonal. However, due to the term which is proportional to the
parameter $\xi$, the 5D bulk mass matrix $C_L$ is not diagonal and
it can be written as,
\begin{eqnarray}\label{CL}
C_L\simeq(\xi V_{5D}\hat{C}_NV^\dagger_{5D}+\hat{C}_e) \; ,
\end{eqnarray}
where $\hat{C}_{N} \equiv d \hat{Y}_{\nu} \hat{Y}_{\nu}^{\dagger}$
and $\hat{C}_{e} \equiv a \hat{Y}_{e} \hat{Y}_{e}^{\dagger}$ are
diagonal. To get Eq.~(\ref{CL}), we have assumed $a\simeq c\simeq d$. The
eigenvalues of $C_L$ give the zero mode localizations of the
$SU(2)_L$ doublets along the fifth dimension. Eq.~(\ref{CL}), which
results from the MFV assumption, leads to a set of conditions that
constrain the 5D bulk mass parameters. 

The non-diagonal term in Eq.~(\ref{CL}) is the source of the FCNCs in the charged lepton sector.
Because this term is  proportional to $\xi$, the size of the contributions to FCNCs is thus determined
by the value of $\xi$, which turns out to be small in order 
to accommodate realistic lepton masses, as we show below.
Because Eq.~(\ref{CL}) involves the unknown mixing matrix $V_{5D}$, to
estimate the value of $\xi$, we take the trace on both sides of
Eq. (\ref{CL})
\begin{eqnarray}
{\rm Tr}(C_L)\simeq c(\xi{\rm Tr}(C_N)+{\rm Tr}(C_e)) \; ,
\end{eqnarray}
where we have replaced $\hat{C}_{N, c}$ with $C_{N, e}$ as trace is
invariant under the basis transformation. To get realistic charged
lepton masses, $c_{L_i}$ and $c_{e_i}$ are in the range of
$(0.4-0.6)$. To accommodate small neutrino masses, the RH neutrinos
must be localized near the Planck brane, and thus $c_{N_i}$ are in
the range of $(1.2-1.5)$. To satisfy the conditions in Eq.
(\ref{CL}), $\xi$ must be in the range of $(0-0.1)$. As a result,
all FCNCs mediated by $(V-A) \times (V+A)$ type operators in the
charged lepton sector are suppressed by a factor of $~{\cal
O}(\xi^2)=(0-0.01)$. For $(V-A)\times (V-A)$ operators, the
suppression factor is ${\cal O}(\xi^4)$. In the limit $\xi=0$, all tree-level 
FCNCs vanish 
and the only sources of flavor violation processes are
the charged current interactions as in the SM\footnote{Similar to the massless neutrino case discussed before, the higher order corrections  involving four or more Yukawa couplings to Eq.~(\ref{eq:mfv2}) could be important. In particular, the term $Y_{e}Y_{e}^{\dagger} Y_{\nu}Y_{\nu}^{\dagger}$ could potentially lead to sizable flavor violation even for vanishing $\xi$. Nevertheless, from the numerical example given below, the Yukawa couplings are smaller compared to 1 and therefore we keep only leading-order terms.}. When $\xi\neq0$,
non-diagonal $C_L$ causes tree-level FCNCs. This is sometimes
referred to as the ``next to minimal flavor violation" (NMFV).

The connection between small neutrino masses and the strong
suppression of FCNCs appears at first to depend crucially on the
assumption that $c \gtrsim d$. As it turns out, small neutrino
masses still implies small $\xi$, even for $c < d$. This is because
by increasing $d$, the 5D Yukawa $Y_{\nu}$ has to decrease to get
the right neutrino masses. In this case, the suppression of FCNCs is
due to the small Yukawa couplings, which is again required by the
smallness of the neutrino masses. Note that in the limit
$Y_{\nu}\rightarrow 0$, there is no FCNCs in the case with MFV
assumption.

Even though the FCNC processes are suppressed due to the MFV
assumption, the MFV assumption does not suppress flavor violation in
the charged currents. In the presence of massive neutrinos, there
are new contributions to the rare decays $\ell_\alpha \rightarrow
\ell_\beta + \gamma$, due to the charged current interactions
involving the exchange of KK W-bosons and KK neutrinos. The zero
mode contributions are suppressed by the GIM mechanism. However, the
contributions of the KK gauge bosons and KK neutrinos invalidate the
conditions required by the GIM mechanism, because of the heavy
masses of the KK neutrinos and the fact that the 4D effective mixing
matrix is not unitary. It has been shown that if the relevant
entries in the 5D Dirac Yukawa matrix are of order unity without the
assumption of MFV, the most stringent constraint, which is from
${\rm BR}(\mu\rightarrow e\gamma)$, requires the first KK mass scale
to be $\gtrsim 25$ TeV, assuming all SM fields are localized on the
TeV brane and only the RH neutrinos are in the bulk ~\cite{kitano}.
In the case with all fermions and gauge bosons in the bulk while
Higgs is localized on the TeV brane, this bound can be relaxed: with
${\cal O}(1)$ Yukawa coupling, the bound on the first KK mode mass
is $\sim 6.7~ {\rm TeV}$, and it can be further relaxed in the bulk
Higgs case\footnote{We thank K. Agashe for pointing this out to
us.}. One way to keep the first KK mass scale accessible to the LHC
while avoiding the $\mu\rightarrow e\gamma$ constraint is to tune
the elements of the 5D Yukawa coupling. We comment that this is
possible with our MFV assumption, which ensures the contributions to
FCNCs are under control. However, in the case with general anarchical
flavor structure in the lepton sector without MFV assumption, the
constraint on the first KK mass cannot be loosened by tuning the
Yukawa couplings, due to the opposite dependence on the Yukawa
couplings in the FCNC contributions.

For simplicity, we do not include CP violation in our numerical results. In this case, the 5D
bulk masses and the leptonic Yukawa sector are determined by only 12
independent physical parameters. This is to be compared with the
generic anarchy case, in which the number of independent parameters
is 27. We show below that with 12 parameters, we are still able to
find solutions that give rise to all realistic lepton masses and
mixing angles, even in the  presence of the constraint given by
Eq.~(\ref{CL}). In the general case with small but not-vanished
$\xi$, $C_L$ is not in the diagonal form. The resulting coupled equations are complicated to solve.
For simplicity, we give a numerical example with $\xi = 0$. In this limit, all FCNCs
vanish. We leave the possibility of having $\xi \ne 0$ for further
investigation.

With $\xi = 0$, all three matrices, $C_L$, $C_e$ and $Y_e$, can be
diagonalized simultaneously. Realistic charged fermion masses arise
with $Y_{e_1}\simeq0.405, \; Y_{e_2}\simeq0.375$ and
$Y_{e_3}\simeq0.354$, assuming $a=c=d=4$. Because the 5D charged
fermion Yukawa matrix is diagonal, the 4D effective PMNS matrix is
determined entirely by  the neutrino sector. In our model, the light
neutrino masses are generated by the Dirac Yukawa couplings,
$m_\nu\simeq vF_LV_{5D}\hat{Y}_\nu F_N$, where the eigenvalues of
$F_{L}$ and $F_{N}$ are given by
\begin{eqnarray}
f_{L_i}=\sqrt{\frac{1-2c_{L_i}}{1-\epsilon^{1-2c_{L_i}}}}, ~~
f_{N_i}=\sqrt{\frac{1-2
c_{N_i}}{1-\epsilon^{1-2c_{N_i}}}}
\end{eqnarray}
and $c_{L_i}$ and $c_{N_i}$ are the eigenvalues of  $C_L$
and $C_N$.

We choose the following 5D parameters as inputs: $
\theta_{12}\simeq1.383,~\theta_{23}\simeq1.358,~\theta_{13}\simeq1.338,
Y_{\nu1}\simeq0.713, ~Y_{\nu2}\simeq0.5634$ and
$Y_{\nu3}\simeq0.5475$, where $\hat{Y}_\nu={\rm
diag}(Y_{\nu_1},Y_{\nu_2},Y_{\nu_3})$ and $\theta_{1,2,3}$ are the
three angles that parametrize the 5D mixing matrix $V_{5D}$.
The 5D Dirac Yukawa coupling matrix is
\begin{eqnarray}\label{5DYnu}
Y_\nu \equiv V_{5D} \hat{Y}_{\nu}
\simeq\begin{pmatrix}0.0307&0.128&0.533\cr-0.275&-0.504&0.123\cr0.657&-0.217&0.0267\end{pmatrix}.
\end{eqnarray}
With these 5D parameters, the effective 4D neutrino oscillation parameters are
\begin{eqnarray}
\nonumber\sin^2\theta^{\nu}_{12}\simeq0.28,~~\sin^2\theta^{\nu}_{23}\simeq0.49,~~\sin^2\theta^{\nu}_{13}\simeq0.023 \; ,
\\
\Delta m^2_{21}\simeq 7.4\times10^{-5}{\rm eV^2},~~\Delta
m^2_{31}\simeq 2.7\times10^{-3}{\rm eV^2} \, ,
\end{eqnarray}
which are in good agreement with experiments within $2\sigma$~\cite{Maltoni:2004ei}. Here we assume a slightly large value of
$d=4$ such that the magnitudes of the 5D Dirac Yukawa couplings are small.
As a consequence, $\mu\rightarrow e\gamma$ mediated by the heavy
neutrinos is suppressed. We estimate the branching fraction ${\rm
Br}(\mu\rightarrow e\gamma)$ to be $\sim10^{-12}$, induced by charged current interactions, 
with the first KK mass scale $\sim
3~{\rm TeV}$.\footnote{In general, there could be UV-sensitive one-loop contributions 
on the IR brane that lead to $\mu-e$ conversion. From our estimate, these contributions are of the order of $\lesssim \mathcal{O}(10^{-20})$ for KK mass $\sim 3$ TeV. Therefore, the bound on the cutoff scale derived from these contributions is less stringent compared to the limit derived from $\mu \rightarrow e \gamma$~\cite{lfvagashe}.} The branching fraction can be further suppresses by tuning the 5D Yukawa couplings and having the Higgs in the bulk.


Even though the eigenvalues of $Y_{\nu}$ are of the same order
$\sim{\cal O}(0.5-0.7)$ and 5D mixing angles are $\sim{\cal O}(1)$,
$Y_{\nu}$ given in Eq.~(\ref{5DYnu}) deviates from the generic
anarchical case with the largest ratio between two elements being
${\cal O}(25)$. This deviation is required to give realistic neutrino
mixing and masses in the case of $\xi\simeq0$, in which all FCNCs 
vanish. The reason is the following: in the generic anarchical case,
the left-handed mixing are given by $V_{ij}\sim f_{L_i}/f_{L_j}$,
and the large solar and atmospheric neutrino mixing angles requires
$f_{L_1}/f_{L_2}\sim1$ and $f_{L_2}/f_{L_3}\sim1$. However, in the
MFV case with $\xi=0$, $f_{L_i}/f_{L_j}$ is fixed by
$\sqrt{m_i/m_j}$ and thus $f_{L_1}/f_{L_2}\simeq0.07$ and
$f_{L_2}/f_{L_3}\simeq0.24$. To  accommodate these ratios as well as
large mixing angles simultaneously, some structure in the 5D Yukawa
couplings is thus needed\footnote{In the generic anarchical case,
one would expect large $\theta^\nu_{13}$ comparable to
$\theta^\nu_{12}$, unless there exists some accidental
cancellation.}. It would be appealing to see if the model can
accommodate $f_{L_1}/f_{L_2}\sim1$ and $f_{L_2}/f_{L_3}\sim1$ in the
presence of small but non-vanishing $\xi$, which may lead to
interesting predictions for tree-level lepton flavor violation
processes.

\section{Conclusions}

We propose a realization of Minimal Flavor Violation in the
lepton sector of the RS model. With the MFV assumption, the only
source of flavor violation are the 5D Yukawa couplings, and the
usual two independent sources of flavor violation are now related.
In the limit of massless neutrinos, there exists a basis
in which the bulk mass matrices and 5D Yukawa matrices are
simultaneously diagonalized, and thus there is no tree level FCNCs.
In the case of massive neutrinos, the contributions to FCNCs in the
charged lepton sector are highly suppressed, as a result of the
smallness of neutrino masses. Even though, the MFV mechanism does not suppress the flavor changing charged-current contributions mediated by KK neutrinos, it nevertheless allows the possibility of tuning the 5D Yukawa couplings to suppress these contributions, which is not possible in the generic anarchical case due to the opposite dependence on the Yukawa couplings between the tree-level FCNC and one-loop charged current contributions. We find that
the first KK mass scale as low as $\sim 3$ TeV can be allowed. In
both cases with either massless or massive neutrinos, we have found
numerical results that give rise to realistic lepton masses and
mixing angles, including those for the neutrinos.

\section{Acknowledgment}
The authors would like to thank Kaustubh Agashe for useful
discussions and comments. This research was supported, in part, by
the National Science Foundation under Grant No. PHY-0709742 at UCI and
PHY05-51164 at KITP. M.-C.C. acknowledges the KITP for its hospitality during the completion of this work.


\begin{thebibliography}{90}
\bibitem{Randall:1999ee}
  L.~Randall and R.~Sundrum,
  Phys.\ Rev.\ Lett.\  {\bf 83}, 3370 (1999).


\bibitem{bulkfields}
  H.~Davoudiasl, J.~L.~Hewett and T.~G.~Rizzo,
  Phys.\ Lett.\  {\bf B473}, 43 (2000);
A.~Pomarol,
  Phys.\ Lett.\ {\bf B486}, 153 (2000);
S.~Chang, J.~Hisano, H.~Nakano, N.~Okada and M.~Yamaguchi,
  Phys.\ Rev. {\bf D62}, 084025 (2000);
S.~J.~Huber and Q.~Shafi,
  Phys.\ Rev. {\bf D63}, 045010 (2001).




\bibitem{bulkfermion1}
  Y.~Grossman and M.~Neubert,
  Phys.\ Lett. {\bf B474}, 361 (2000);
  S.~J.~Huber and Q.~Shafi,
  {\it ibid.} {\bf 498}, 256 (2001);
  S.~J.~Huber and Q.~Shafi,
  {\it ibid.} {\bf 512}, 365 (2001).

\bibitem{bulkfermion2}
  T.~Gherghetta and A.~Pomarol,
  Nucl.\ Phys. {\bf B586}, 141 (2000).





\bibitem{Chen:2005mz}
  S.~J.~Huber and Q.~Shafi,
  Phys.\ Lett.\  {\bf B544}, 295 (2002);
  M.-C.~Chen,
  Phys.\ Rev.\  {\bf D71}, 113010 (2005).

\bibitem{custodial}
  K.~Agashe, A.~Delgado, M.~J.~May and R.~Sundrum,
  JHEP {\bf 0308}, 050 (2003).

\bibitem{branekinetic}
  H.~Davoudiasl, J.~L.~Hewett and T.~G.~Rizzo,
  Phys.\ Rev.\  {\bf D68}, 045002 (2003);
  M.~S.~Carena, E.~Ponton, T.~M.~P.~Tait and C.~E.~M.~Wagner,
  {\it ibid.} {\bf  67}, 096006 (2003);
  M.~S.~Carena, A.~Delgado, E.~Ponton, T.~M.~P.~Tait and C.~E.~M.~Wagner,
  {\it ibid.} {\bf 71}, 015010 (2005).


\bibitem{rsgim1}
  K.~Agashe, G.~Perez and A.~Soni,
  Phys.\ Rev.\ Lett.\  {\bf 93}, 201804 (2004);
  K.~Agashe, G.~Perez and A.~Soni,
  Phys.\ Rev.\  {\bf D71}, 016002 (2005);
  K.~Agashe, M.~Papucci, G.~Perez and D.~Pirjol,
  arXiv:hep-ph/0509117;
  Z.~Ligeti, M.~Papucci and G.~Perez,
  Phys.\ Rev.\ Lett.\  {\bf 97}, 101801 (2006).




\bibitem{Bona:2007vi}
  M.~Bona {\it et al.}  [UTfit Collaboration],
  JHEP {\bf 0803}, 049 (2008);
K.~Agashe {\it et al.},
  Phys.\ Rev.\  {\bf D76}, 115015 (2007).





\bibitem{Csaki:2008zd}
  C.~Csaki, A.~Falkowski and A.~Weiler,
  arXiv:0804.1954 [hep-ph].



\bibitem{kitano}
  R.~Kitano,
  Phys.\ Lett.\  {\bf B481}, 39 (2000).

\bibitem{lfv}
   S.~J.~Huber,
  Nucl.\ Phys.\   {\bf B666}, 269 (2003);
  G.~Moreau and J.~I.~Silva-Marcos,
  JHEP {\bf 0603}, 090 (2006).
  S.~Davidson, G.~Isidori and S.~Uhlig,
  Phys.\ Lett.  {\bf B663}, 73 (2008).
  
\bibitem{lfvagashe}
  K.~Agashe, A.~E.~Blechman and F.~Petriello,
  Phys.\ Rev.\  {\bf D74}, 053011 (2006);

\bibitem{Cacciapaglia:2007fw}
  G.~Cacciapaglia, C.~Csaki, J.~Galloway, G.~Marandella, J.~Terning and A.~Weiler,
  JHEP {\bf 0804}, 006 (2008).

\bibitem{rsmfv}
  A.~L.~Fitzpatrick, G.~Perez and L.~Randall,
  arXiv:0710.1869 [hep-ph].




\bibitem{mfv}
  E.~Gabrielli and G.~F.~Giudice,
  Nucl.\ Phys.\   {\bf B433}, 3 (1995),
  [Erratum-ibid.\ {\bf 507}, 549 (1997)];
  A.~Ali and D.~London,
  Eur.\ Phys.\ J.\   {\bf C9}, 687 (1999);
  A.~J.~Buras, P.~Gambino, M.~Gorbahn, S.~Jager and L.~Silvestrini,
  Phys.\ Lett.\  {\bf B500}, 161 (2001);
  G.~D'Ambrosio, G.~F.~Giudice, G.~Isidori and A.~Strumia,
  Nucl.\ Phys.\   {\bf B645}, 155 (2002);
 V.~Cirigliano, B.~Grinstein, G.~Isidori and M.~B.~Wise,
  Nucl.\ Phys.\  {\bf B728}, 121 (2005).


\bibitem{collider}
  H.~Davoudiasl, J.~L.~Hewett and T.~G.~Rizzo,
  Phys.\ Rev.\  {\bf D63}, 075004 (2001);
K.~Agashe, G.~Perez and A.~Soni,
  {\it ibid.}  {\bf D75}, 015002 (2007);
  K.~Agashe, A.~Belyaev, T.~Krupovnickas, G.~Perez and J.~Virzi,
  {\it ibid.}  {\bf 77}, 015003 (2008);
 R.~Contino, T.~Kramer, M.~Son and R.~Sundrum,
  JHEP {\bf 0705}, 074 (2007);
B.~Lillie, L.~Randall and L.~T.~Wang,
  {\it ibid.} {\bf 0709}, 074 (2007);
A.~L.~Fitzpatrick, J.~Kaplan, L.~Randall and L.~T.~Wang,
  {\it ibid.} {\bf 0709}, 013 (2007);
K.~Agashe, H.~Davoudiasl, G.~Perez and A.~Soni,
  Phys.\ Rev.\   {\bf D76}, 036006 (2007);
M.~Arai, N.~Okada, K.~Smolek and V.~Simak,
  {\it ibid.}   {\bf D75}, 095008 (2007);
C.~Dennis, M.~Karagoz Unel, G.~Servant and J.~Tseng,
  arXiv:hep-ph/0701158;
  F.~Ledroit, G.~Moreau and J.~Morel,
  JHEP {\bf 0709} (2007) 071;
P.~J.~Fox, Z.~Ligeti, M.~Papucci, G.~Perez and M.~D.~Schwartz,
  arXiv:0704.1482;
      A.~Djouadi, G.~Moreau and R.~K.~Singh,
  Nucl.\ Phys.\  {\bf B797}, 1 (2008);
    B.~Lillie, J.~Shu and T.~M.~P.~Tait,
  Phys.\ Rev.\  {\bf D76}, 115016 (2007);
  H.~Davoudiasl, G.~Perez and A.~Soni,
  arXiv:0802.0203 [hep-ph];
  U.~Baur and L.~H.~Orr,
  arXiv:0803.1160.

\bibitem{bulkhiggs}
  K.~Agashe, R.~Contino and A.~Pomarol,
  Nucl.\ Phys.\  {\bf B719}, 165 (2005);
  H.~Davoudiasl, B.~Lillie and T.~G.~Rizzo,
  JHEP {\bf 0608} (2006) 042.


\bibitem{Maltoni:2004ei}
  M.~Maltoni, T.~Schwetz, M.~A.~Tortola and J.~W.~F.~Valle,
  New J.\ Phys.\  {\bf 6}, 122 (2004).

\end{thebibliography}
\end{document}